\documentclass[aps,prl,twocolumn,showpacs,superscriptaddress]{revtex4}

\usepackage{epsfig}
\usepackage{amsmath}
\usepackage{amssymb}
\usepackage{textcomp}

\begin{document}
\title{Glassy Dislocation Dynamics in 2-D Colloidal Dimer Crystals}
\author{Sharon J. Gerbode}
\affiliation{Department of Physics, Cornell University, Ithaca NY 14853}
\author{Umang Agarwal}
\affiliation{Chemical and Biomolecular Engineering, Cornell University, Ithaca NY 14853}
\author{Desmond C. Ong}
\affiliation{Department of Physics, Cornell University, Ithaca NY 14853}
\author{Chekesha M. Liddell}
\affiliation{Materials Science and Engineering, Cornell University, Ithaca NY 14853}
\author{Fernando Escobedo}
\affiliation{Chemical and Biomolecular Engineering, Cornell University, Ithaca NY 14853}
\author{Itai Cohen}
\affiliation{Department of Physics, Cornell University, Ithaca NY 14853}

\begin{abstract}
\label{sec:abstract} Although glassy relaxation is typically associated with disorder, here we report on a new type of glassy dynamics relating to dislocations within 2-D crystals of colloidal dimers. Previous studies have demonstrated that dislocation motion in dimer crystals is restricted by certain particle orientations. Here, we drag an optically trapped particle through such dimer crystals, creating dislocations. We find a two-stage relaxation response where initially dislocations glide until encountering particles that cage their motion. Subsequent relaxation occurs logarithmically slowly through a second process where dislocations hop between caged configurations. Finally, in simulations of sheared dimer crystals, the dislocation mean squared displacement displays a caging plateau typical of glassy dynamics. Together, these results reveal a novel glassy system within a colloidal crystal.
\end{abstract}


\maketitle \setcounter{page}{1} \thispagestyle{empty}

\begin{figure}[ht]
\centering
\includegraphics[height=4.8in]{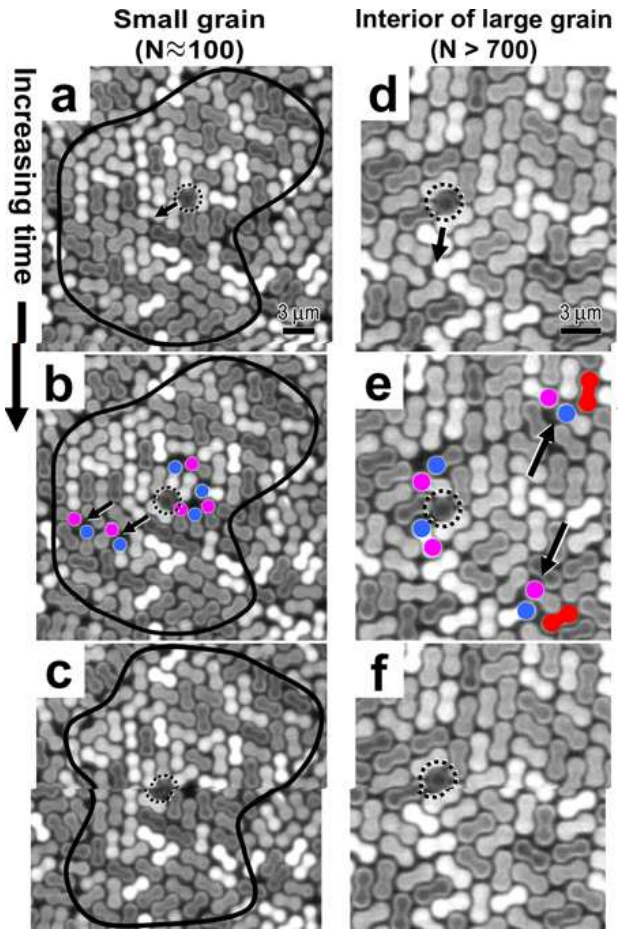}
\caption{Micrographs illustrating a local perturbation experiment. Five-fold and seven-fold coordinated defects created by the perturbation are marked with pink and blue dots, respectively. (a) A single spherical intruder particle in a small grain is dragged by 1 LC using optical tweezers. (b) Two dislocations marked by arrows glide to the grain edge and are absorbed by the grain boundary (solid closed curve).  (c) The intruder particle remains stationary after the tweezers are turned off. The vacancy left behind has no topological charge and is stable. (d-f)~In a large grain (only the central part is shown here) the dislocations marked by arrows are obstructed by the particles in red.  When the tweezers are turned off, the blocked dislocations return along their glide paths, recombining and causing the intruder particle to recoil.}
\label{fig:dragPics}
\end{figure}

Dislocation mobility is central to both the materials properties and relaxation mechanisms of crystalline materials \cite{Orowan1934,Polanyi1934,Taylor1934}. Previous studies of dislocation motion in colloidal crystals composed of spherical particles have allowed a particle-scale view of defect formation and transport \cite{Schall2004,LibalReichhardt2007,Zahn1999,Lekkerkerker2005}. More recent experiments \cite{Gerbode2008, Lee2008} have begun to interrogate the role of particle anisotropy in determining the rules of defect motion.  In particular, studies of dislocation motion in crystals of colloidal dimer particles have uncovered novel restrictions on dislocation mobility.  In these ``degenerate crystals'', the dimer lobes occupy triangular lattice sites while the particles are randomly oriented among the three crystalline directions, as shown in Fig.~\ref{fig:dragPics}. One consequence of the random orientations of the dimers is that dislocation glide is severely limited by certain particle arrangements in degenerate crystals \cite{Gerbode2008}. 

The present work utilizes local mechanical perturbation experiments to investigate the effects of this restricted dislocation motion on the relaxation mechanisms in degenerate crystals of colloidal dimers.  Holographic optical tweezers are used to manipulate single lobe-sized spherical intruder particles within an otherwise pure degenerate crystal grain, deforming the crystal and introducing defects. During the subsequent relaxation of the degenerate crystal, dislocations formed during the deformation leave the grain, either via annihilation with other dislocations or by moving to a grain boundary. Interestingly, we find that in large crystalline grains this dislocation relaxation occurs through a two-stage process reminiscent of slow relaxations in glassy systems, suggesting the novel concept that glassy phenomena may be present within certain kinds of colloidal crystals.

We synthesize sterically stabilized silica dimer shell particles with spherical lobes of diameter $1.3~\mu$m and lobe separation $1.4~\mu$m as previously described \cite{Gerbode2008, Lee2008}.  For the spherical intruder particles, we include 1\% volume fraction of $1.3~\mu$m polystyrene spherical particles that are coated with silica and sterically stabilized using polyvinylpyrrolidone (PVP) to produce surface chemistry identical to the dimer particles.  All particles are suspended in an aqueous solution of dimethylsulfoxide (DMSO) that index-matches the silica shells (with density mismatch $\sim$~0.5~g/mL). The polystyrene cores of the spherical intruder particles remain index-mismatched, allowing for optical manipulation using laser tweezers.  Thus, perturbations to the degenerate crystal are applied only via the motion of the intruder particle. The suspension is pipetted into a sealed wedge-shaped glass cell as previously described \cite{Gerbode2008} and the particle area fraction (maintained at $\sim$~0.8) is controlled by tilting the cell so that particles sediment into the viewing region, which only accommodates a 2-D monolayer of particles. During the perturbation experiments, the crystal is imaged using a confocal microscope integrated with the holographic optical tweezer system (Arryx, Inc) \cite{Dufresne1998}.

In preparation for a local perturbation experiment, one intruder particle is moved to the center of a degenerate crystal grain, and then the system is allowed to relax for at least 10 hours until the grain is defect-free. Fig.~\ref{fig:dragPics}a-c shows a typical drag experiment for a small degenerate crystal grain consisting of $N \approx 100$ dimer particles. At time $t=0$~s  (Fig.~\ref{fig:dragPics}a) the optical tweezers are turned on and used to drag the intruder particle by one lattice constant (LC) along a lattice direction, arriving at the new lattice position at $t=20$~s.  Then the optical tweezers are turned off and the grain is imaged until all the defects leave the grain or recombine (Fig.~\ref{fig:dragPics}c). Dislocations created by the drag deformation are identified from the microscope images by calculating the number of nearest neighbors for each lobe using Voronoi analysis. Each dislocation, consisting of paired five-fold and seven-fold coordinated lobes, is tracked over the experiment.

We observe different relaxation behavior depending on the degenerate crystal grain size. In small grains, the intruder particle remains stationary after being dragged to its new lattice position, even after the tweezers are turned off.  In such cases, dislocations like those identified by arrows in Fig.~\ref{fig:dragPics}b glide to the edge of the grain and are absorbed by the grain boundary, which we identify using Voronoi analysis (details in the supplementary materials). In larger grains (Fig.~\ref{fig:dragPics}d-f), we observe that dislocations gliding toward the grain boundary are more likely to encounter glide-blocking particles whose orientations prevent further glide \cite{Gerbode2008}. For example, the dislocations marked by arrows in Fig.~\ref{fig:dragPics}e must travel so far to reach the grain edge (beyond the field of view of Fig.~\ref{fig:dragPics}e) that they encounter the blocking particles shown in red. Consequently, when the optical tweezers are turned off, these dislocations return toward the intruder particle, recombining with the other defects produced by the deformation, and causing the intruder particle to recoil.  

\begin{figure}[hb]
\centering
\includegraphics[height=2.0in]{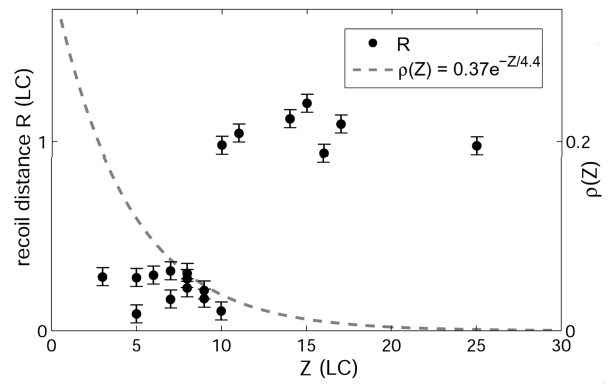}
\caption{Recoil distance $R$ as a function of glide distance $Z$ to the nearest grain boundary.  The error bars represent uncertainty in particle position. The distribution of unrestricted glide distances $\rho(Z)$ (dashed gray curve) is reprinted from \cite{Gerbode2008}.}
\label{fig:recoilPlot}
\end{figure}

To elucidate the effect of grain size on the recoil distance of the intruder particle we conduct 19 independent deformation experiments on degenerate crystal grains of varying size. We find that a crossover from the stationary response to the recoiling response occurs when dislocations produced during the deformation must traverse distances $Z > 10$~LC to reach a grain boundary (Fig.~\ref{fig:recoilPlot}). This result is consistent with the previously reported distribution of unrestricted dislocation glide distances $\rho(Z)$, which was found to decay exponentially (dashed gray curve in Fig.~\ref{fig:recoilPlot}) \cite{Gerbode2008}. Together these results depict the probabilistic process, governed by the distribution $\rho(Z)$, by which a deformation-induced dislocation may glide to a grain edge without meeting an obstacle.

Although unrestricted dislocation glide over long distances is improbable, multi-defect mechanisms such as dislocation reactions can allow dislocations to bypass glide-blocking obstacles and achieve long range motion \cite{Gerbode2008}. However, since the energy required for such dislocation interactions is higher than that for simple dislocation glide \cite{Gerbode2008}, larger mechanical deformations are required to access such processes.  Furthermore, mechanisms involving more than one defect are statistically less probable, and consequently may require longer waiting times. 

In order to probe such multi-defect relaxation responses we conduct long time relaxation experiments where a perturbation is rapidly applied to a large degenerate crystal grain. Before each experiment, one intruder particle is placed near the center of a large degenerate crystal grain containing more than 700 dimer particles. At time $t=0$~s, the intruder particle is quickly dragged by 3 LC, arriving at its new lattice position at $t=5$~s.  The optical trap is turned off and the system is imaged for 10~hr, or until all the defects produced by the deformation leave the crystal grain. We record the number of defects by counting the lobes within the crystal grain that do not have six nearest neighbors.  For example, a dislocation consisting of one five-fold and one seven-fold coordinated lobe is counted as two defects.

To compare the relaxation response of degenerate crystals with a system where dislocations are known to glide without geometric restrictions, we repeat this experimental procedure on crystals of spheres. Here we prepare crystals of PVP-stabilized 1~$\mu$m silica spheres suspended in an index-matching aqueous solution of DMSO, and we use 1~$\mu$m PVP-stabilized silica-coated polystyrene spheres as the intruder particles.

\begin{figure}[h]
\centering
\includegraphics[height=2.3in]{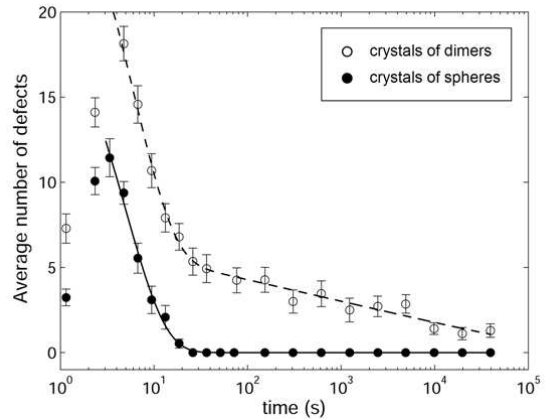}
\caption{Average number of defects versus time for large ($>700$ particles) degenerate crystals and crystals of spheres.  Error bars represent the standard error of the mean. The solid line is the best fit exponential for the decay in the sphere data ($R^2=0.99$); the dashed line is the best fit sum of an exponential and a logarithmic decay for the dimer data ($R^2=0.99$).}
\label{fig:defectDecay}
\end{figure}

The results from long time relaxation experiments on both degenerate crystals and crystals of spheres are shown in Fig.~\ref{fig:defectDecay}. In crystals of spheres, the decrease in the average number of defects $\overline{N_d}$ after the optical trap is turned off follows an exponential decay.  The best fit exponential for this decay is $\overline{N_d} = 23e^{-t/\tau_S}$, where $\tau_S=5 \pm 0.5$~s. In degenerate crystals, the initial decay of $\overline{N_d}$ is also characterized by an exponential decay, but a long tail is evident for later times.  The best fit curve for this system combines both exponential and logarithmic decay terms: $\overline{N_d} = 25e^{-t/\tau_\beta}+0.55 \ln(1+\frac{\tau_\alpha}{t})$, where $\tau_\beta= 6 \pm 1$~s and $\tau_\alpha=2 \pm 1\times10^5$~s.  These data depict strikingly different relaxation responses in the two systems.  In crystals of spheres, dislocations leave the grain via one fast mechanism with characteristic timescale $\tau_S$.  In degenerate crystals, a similar initial fast response with timescale $\tau_\beta$ is followed by a much slower process with a timescale $\tau_\alpha$ that is 5 orders of magnitude larger than $\tau_\beta$. 

Such two-stage relaxation is reminiscent of the slow relaxation dynamics common to many glassy systems \cite{Sjogren1989,Granular2004,Weeks2002}.  Studies of colloidal glasses have revealed a caging effect where particles are confined by their neighbors \cite{Weeks2002} and are transported through the glass in two stages: an initial fast diffusion until encountering the caging neighbors, and a much slower cage-hopping process caused by multi-particle rearrangements that shift the cage structure. This two-stage glassy relaxation is characterized by a long tail similar to that observed in defect relaxation for degenerate crystals (Fig.~\ref{fig:defectDecay}), suggesting that defect dynamics in dimer crystals is glassy. 

Similarly to a particle in a glass, a dislocation in a degenerate crystal is caged. Initially, it moves via unrestricted glide at a rate characterized by $\tau_\beta$, until it reaches particles with glide-blocking orientations. Here it remains caged until a second process characterized by the time scale $\tau_\alpha$ can allow it to hop to a new region of unrestricted glide. This two stage relaxation process also explains the recoil measurements in Fig.~\ref{fig:recoilPlot}. For small grains where dislocations only need to travel a short distance to reach the grain boundary, relaxation can occur through unrestricted glide alone. For larger grains, both stages of relaxation would be needed for the dislocations to exit the grain. However, this would require sufficiently long time scales for cage hopping processes to occur. Particles in the recoil experiments are only dragged 1 LC with a drag duration of $\tau_\beta <$ 20~s $\ll \tau_\alpha$, which does not create enough defects to trap the system and allow enough time for multidefect relaxation processes, thus leading to the observed recoil in large grains. 

\begin{figure}[ht]
\centering
\includegraphics[height=2.6in]{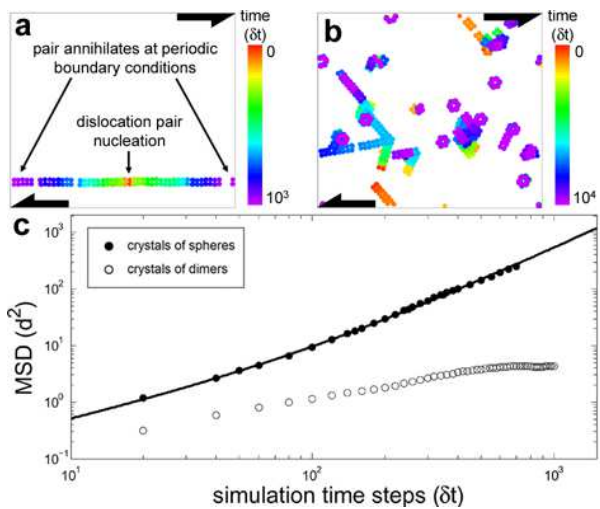}
\caption{(Color online) Defect trajectories in simulated crystals under shear (large arrows). (a) Dislocations in a crystal of spheres glide along straight paths in response to the applied shear. (b) Dislocations in a degenerate crystal follow crooked trajectories as they hop between caged configurations. (c) MSD of individual dislocations in crystals of spheres and degenerate crystals. Error bars are smaller than the markers. The solid curve is the best fit for \emph{MSD}~$= \langle \Delta r^2 \rangle = u^2 {t}^2 + 2Dt$.}
\label{fig:simulation}
\end{figure}

Glassy relaxation can be directly observed in colloidal glasses by measuring the particle mean square displacement (MSD) versus time. Visualizing dislocation trajectories in local deformation experiments is complicated by the fact that all the defects originate near the intruder particle, rather than uniformly throughout the system as would occur during macroscopic deformations. To explore such macroscopic deformations we run Non Equilibrium Molecular Dynamics (NEMD) simulations of dimer and sphere crystals under uniform shear. This approach probes defect dynamics in a non-equilibrium ``steady state'' rather than a perturbation. Furthermore, such simulations allow for comparison of dimer and sphere crystals in ideal conditions without grain boundaries or particle polydispersity. The simulations are conducted in a 2-D canonical ensemble wherein hard spheres (or dimer lobes) interact via the Weeks-Chandler-Andersen potential \cite{WCA1971}.  In reduced Lennard-Jones units, the system has size $50~\times~50$, sphere or lobe diameter $d = 1$, dimer lobe separation 1.07 maintained using holonomic constraints, and temperature of 1 maintained via a configurational thermostat \cite{Travis2005}.  Area fractions of 0.808 for dimers and 0.792 for spheres are chosen to have both systems at identical pressure, temperature, and chemical potential.  A homogeneous shear field is applied to initially defect-free crystals using SLLOD equations of motion \cite{Evans1990} and Lees-Edwards periodic boundary conditions \cite{LeesEdwards1972}.  The equations of motion are integrated using the Runge Kutta 4th order method with time step $\delta_t = 0.01$.  Uniform shear is applied at a strain rate $\dot{\gamma} = 5\times10^{-5}$ per $\delta t$, a value small enough to drive dislocation motion at a speed independent of $\dot{\gamma}$. Dislocations are monitored after the number of defects has reached a steady value (full simulation details provided in the supplementary materials section, below).  

These simulations confirm strikingly different defect transport mechanisms in the two systems. Fig.~\ref{fig:simulation}a-b show plots of defect positions, where color corresponds to time. In crystals of spheres, dislocation pairs form and glide apart along straight paths nearly aligned with the shear direction (Fig.~\ref{fig:simulation}a). In stark contrast, defects in degenerate crystals follow crooked trajectories (Fig.~\ref{fig:simulation}b) and do not glide throughout the crystal, but rather are trapped within local cages. This caging is evident in the dislocation MSD, shown for both degenerate crystals and crystals of spheres in Fig.~\ref{fig:simulation}c.  To calculate the dislocation MSD versus time, individual dislocations that can be resolved from other nearby defects are tracked between time steps. In crystals of spheres, dislocation motion is consistent with a biased 1D random walk (solid curve, Fig.~\ref{fig:simulation}c) with a diffusion constant $D = 2.3 \pm 0.2 \times 10^{-2}~d^2/\delta t$, and drift speed $u = 2.23 \pm 0.05 \times 10^{-2}~d/\delta t$. The dislocation MSD in degenerate crystals displays the characteristic plateau typical of particles in glassy systems.

In conclusion, we have uncovered a novel glassy system where the constituent particles are assembled into an ordered crystalline structure, but the dislocations within these crystals are caged and demonstrate slow, two-stage glassy relaxation. Such crystals also have a natural mechanism for encoding memory effects commonly observed in glasses, since moving dislocations reorient the dimer particles, encoding constraints for subsequent defect motion. While the transition into the glassy state typically occurs through an increase in particle density \cite{LiuNagel1998}, here the glassy dislocation dynamics arises from constraints in an already dense crystal. Future studies of crystals of spheres doped with increasing concentrations of dimers will probe for this transition and determine whether it can be thought of more generically as an additional route to the jammed state. Finally, unlike typical jammed systems, increasing the dislocation density can reduce the amount of caging, suggesting that degenerate crystals may be self-healing materials.

We thank R. Ganapathy, X. Cheng, and the A. Liu group for helpful discussion.  This work was supported by the DOE, Basic Energy Sciences, Grant \#ER46517.

\newpage
\section{Glassy Dislocation Dynamics in 2-D Colloidal Dimer Crystals: Supplementary Materials}

\section{Molecular Dynamics Simulations}
We have performed 2-D Number Volume Temperature (NVT) Non Equilibrium Molecular Dynamics (NEMD) simulations of both pure sphere and pure dimer crystals.  The separation between lobes in the dimer particles is set to be $1.07$ times the diameter of the spherical lobes, in approximation of the experimental system. The hard spheres (or dimer lobes) interact with each other via the Weeks-Chandler-Andersen potential \cite{WCA1971} given by

 \[U(r) = \left\{
\begin{array}{l l}
  4(r^{-12}-r^{-6})+1 & \quad \mbox{$r \leq r_m=2^\frac{1}{6}$}\\
  0 & \quad \mbox{$r > r_m$}\\ \end{array} \right. \]

\noindent where r is the interparticle distance and $r_m$ is the minimum of the potential. In this and all the following equations, all the properties are expressed in terms of reduced Lennard-Jones units. 

The crystals are initially defect-free, with every sphere (or dimer lobe) located exactly on a triangular lattice site. Dimer crystals are in a degenerate crystalline state such that the dimers are oriented randomly along any of the three lattice directions \cite{Woj1991}. The simulations of both dimer and sphere crystals are performed at equivalent area fractions, as defined below, at reduced temperatures $T^* = 1.0$ (sufficiently below their melting temperatures) with a time step $\delta t = 10^{-2}$. To preserve the heterogeneities in the system we use a novel configurational thermostat which does not constrain the flow profile \cite{Travis2005}.

\subsection{Area fraction}
In order to fairly compare crystals of dimers with crystals of spheres at equivalent area fractions, we first perform direct interfacial NVT simulations of compound systems of sphere and dimer lattices. On equilibration, once both lattices share the same chemical potential, pressure and temperature, these simulations provide estimates of the coexistence area fractions for the two species. Then, for our NEMD simulations, we have used one set of these equivalent area fractions: 0.808 for dimers and 0.792 for spheres, both well above the melting densities.  

\begin{figure}[ht]
\centering
\includegraphics[height=2.2in]{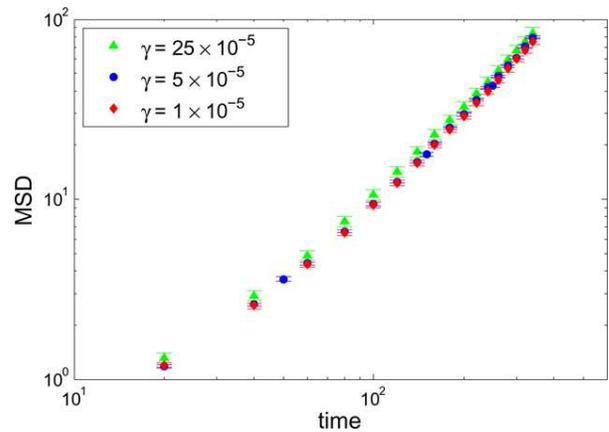}
\caption{Dislocation mean squared displacement in crystals of spheres under uniform shear at three different shear rates $\dot{\gamma}$. Error bars represent the standard error of the mean.}
\label{fig:shearIndep}
\end{figure}

\subsection{Shear rate}
Both the crystals of pure spheres and the crystals of pure dimers are sheared at a shear rate $\dot{\gamma} = 5\times10^{-5}$.  In order to choose the value of $\dot{\gamma}$, several factors were considered.  First, in order to study dislocation motion in the crystalline regime, the shear rate must not be high enough to destroy the crystalline order. When shear is applied to the initially defect-free system, some defects are created, thereby reducing the crystalline order.  However, if the shear rate is chosen to be low enough, then after some characteristic `stabilization time' -- about $10^6$ time steps in our simulations -- the number of defects in the system fluctuates around a steady value and the crystalline order does not decrease further. 

A second important consideration is the effect of shear rate on dislocation glide speed in crystals of spheres. To ensure that we are studying dislocation motion in a `quasi-static' regime where dislocation motion is independent of the applied shear rate, we have also tracked dislocation motion in crystals sheared at two additional shear rates: $\dot{\gamma} = 25\times10^{-5}$ and $\dot{\gamma} = 1\times10^{-5}$. We compare the dislocation mean squared displacement (MSD) for the 3 shear rates and find that while dislocation motion is slightly faster for $\dot{\gamma} = 25\times10^{-5}$, the lower two shear rates are indistinguishable within the error bars (Fig.~\ref{fig:shearIndep}).  This indicates that at $\dot{\gamma} = 5\times10^{-5}$, we are in the regime where dislocation motion is independent of $\dot{\gamma}$.

This shear rate independence can be understood by considering dislocation motion in crystals of spheres, where we find that long periods of dislocation inactivity are punctuated by brief glide events.  In such glide events, two dislocations form and glide apart along the lattice axis aligned with the shear direction. Between the glide events, shear stress caused by the applied shear rate $\dot{\gamma}$ slowly builds up until enough stress is present to separate a pair of dislocations. Once the dislocations start to separate, they quickly glide apart and annihilate after traversing the periodic boundary conditions.  Changing the shear rate $\dot{\gamma}$ only affects the average waiting time between glide events, not the speed of dislocation motion during the events.  To quantify how small the applied shear rate is, we compare the time $\tau_{\dot{\gamma}}$ for the applied shear to cause a shear displacement of 1~LC to the time $\tau_{d}$ for a dislocation pair to separate across the simulation box, causing a shear displacement of 1~LC.  For $\dot{\gamma} = 5\times10^{-5}$, $\tau_{\dot{\gamma}} = 4 \times 10^4$, while $\tau_{d} = 990 \pm 16$ (mean $\pm$ standard error of the mean).  Thus, $\tau_{\dot{\gamma}}$ is about 40 times larger than $\tau_{d}$.

\subsection{Calculating dislocation mean squared displacement}
To calculate the dislocation MSD, we have tracked the positions of individual dislocations consisting of paired 5-fold and 7-fold coordinated lobes (or spheres).  In both degenerate crystals of dimers and crystals of spheres, our analysis is restricted to dislocations that can be distinctly resolved from any other nearby defects. In dimer crystals, this requirement limits our analysis to dislocation motion before or after -- but not during -- a dislocation reaction.

In order to correct for affine displacements imposed by the applied external shear $\dot{\gamma}$, relative dislocation $x$-displacements $\Delta x ( \tau )$ were shifted by the accumulated strain displacement $\delta x ( \tau )$ corresponding to the relative $y$-displacement $\Delta y ( \tau )$.  The shear-corrected dislocation displacement $\Delta x_{c} ( \tau )$ after a time interval $\tau$ is given by:
\begin{eqnarray*}
  \Delta x_c (\tau ) & = & \Delta x (\tau) - \delta x (\tau ) \\
  & = & \Delta x (\tau ) - \dot{\gamma} \tau \Delta y (\tau )
\end{eqnarray*}
Typical values of $\delta x ( \tau )$ for time intervals $\tau$ measured in calculating the dislocation MSD were approximately 3 orders of magnitude smaller than $\Delta x_{c} ( \tau )$.

\begin{figure}[t]
\centering
\includegraphics[height=2.3in]{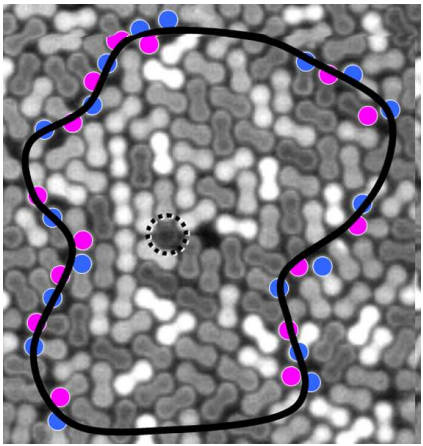}
\caption{The grain boundary defects are highlighted around the small grain also shown in Fig.~1c of the manuscript.  Pink and blue dots represent lobes having a number of nearest neighbors equal to 5 or 7, respectively.  The dotted circle marks the spherical intruder particle.}
\label{fig:grainBoundary}
\end{figure}

\section{Experimental Techniques}
\subsection{Computing grain boundaries}
To locate the grain boundaries in the experimental degenerate crystals, Voronoi analysis is used to compute the number of nearest neighbors for each lobe.  Crystalline grains containing many ordered lobes with 6 nearest neighbors each are surrounded by a string-like collection of defect lobes having a number of nearest neighbors not equal to 6.  The grain boundary defects for the small grain shown in Fig.~1c of the manuscript are explicitly highlighted in Fig.~\ref{fig:grainBoundary}.

\bibliography{GerbodeRefs}

\end{document}